# Enhancement in absorption of amorphous Si solar cell by using conducting anti-reflection coating and plasmonic nano-structured back reflector

Sandipta Roy,[a,*] Arnab Pattanayak,[a] Siddartha P. Duttagupta,[a,b]

[a]Centre for Research for nanotechnology and Science, Indian Institute of Technology Bombay, Mumbai-4000046, India

[b]Department of Electrical Engineering, Indian Institute of Technology Bombay, Mumbai-4000046, India

**Abstract**. In this work is comprising of simulation results of amorphous Si (a-Si) solar cell, in which ITO was used as an anti-reflection coating (ARC). The thickness of ITO was optimized (65 nm) to achieve better performance for spectral region of 350-800 nm. The reflectivity of the solar cell using ITO and SiNx as an ARC were compared. The result is found similar. Further to enhance the light absorption in a-Si, silver nano hemispherical back reflector was used. And it was found that the average absorption by the a-Si has been improved to ~80% by using ITO and hemi spherical nano structured back reflector.

*Sandipta Roy**, E-mail: sandipta.r@iitb.ac.in

## 1 Introduction

Due to the requirement of alternative energy source in near future, an intense investigation has been made towards the solar energy harvesting, where the solar energy is converted into electricity by means of electron hole pair generation in semiconductor device. Amorphous silicon (a-Si) is one of the promising photovoltaic (solar) material used for solar cell fabrication due to its nontoxicity, abundance, mature processing technology. Additionally, the effective absorption depth of a-Si is only about 1 μm, i.e., 2 orders of magnitude thinner than that of crystalline silicon, which favors to fabrication of thin film solar cell.

The common factor which affects the efficiency of a solar cell is the reflection loss from the front side of the device. The optical impedance mismatch (due to the difference of refractive indices, r.i.) of air and a-Si is primarily responsible for such behavior. One of the well-known technique to reduce the reflection loss is surface texturing, but this cost to high surface recombination centre of the minority carriers [1,2]. Therefore, the alternative technique, like quarter wavelength anti-reflection coating (ARC) is the most favorable solution and widely used. Over the years several materials have been demonstrated as an ARC; like: $TiO_2$ (r.i.= 2.3), $Si_3N_4$ (r.i. = 1.9), $Al_2O_3$ (r.i.= 1.8–1.9), $SiO_2$ (r.i.= 1.4–1.5), $Ta_2O_5$ (r.i.= 2.1–2.3) [3,4]. Among them silicon nitride ($SiN_x$) is the common material used at present for silicon solar cell [5]. The ARC must be transparent to reduce reflection and must be highly conductive to serve as a front contact for the application in thin film solar cells. Therefore, in this work we are using indium tin oxide (ITO, transparent and conducting) as an ARC materials, which can serve as top electrode contact for a-Si thin film solar cell as well. Furthermore, a back reflector was used to minimize the transmission loss of the solar cell.

Silicon is a poor absorber to the higher wavelength (greater than 700 nm) region of solar spectrum. A thick semiconducting layer is required to trap that region, but this may reduce the efficiency of photo generated carrier collection (the minority carrier diffusion length in the a-Si is ~300 nm [4]). Thinning of active material thickness may solve the problem, but this will cost to the effective light absorption in the active region. Efficient light trapping along with effective carrier collection is the key requirement to realize the efficient performance of solar cell. The promising solution to trap the light effectively is incorporation of nano structure in the device [6,7]. The nano structures scatter the light in the active materials and increase the path length in it. And it is well known that if the scattering angle inside the a-Si become more than 16°, the light will be trap inside the active region (a-Si)[8] and thus the performance of solar cell will improve.

In this work, we are presenting simulation results of optimization ITO thickness ($d_{ITO}$) as an efficient antireflection coating on a-Si. And further the simulation was done with the nano structured back reflector to study the enhancement of the light trapping in the solar cell. The simulation was performed by using CST Microwave Studio.

## 2 Design and Simulation

*2.1 ITO as antireflection coating*

The proposed structure of a-Si solar cell with the anti-reflection coating and a back reflector is shown in the Fig. 1 (a-c) (a 3D schematic is shown in Fig. 9(b)). To understand the effect of the ARC we have performed full wave simulation of a bare a-Si/back metal and ITO/a-Si/back metal structures. The performance of the device with ITO coated was compared with the standard 60 nm SiN coated sample (simulated results of SiN/a-Si/back reflector). The thickness of the ITO was varied from 40 to 105 nm to optimize at minimum average reflection form the structure in the spectral region of interest (350–800 nm). The thickness of the a-Si is considered to be 250 nm and silver was used as a back reflector and nano particle material.

*2.2 Plasmonic light trapping in thin film solar cell*

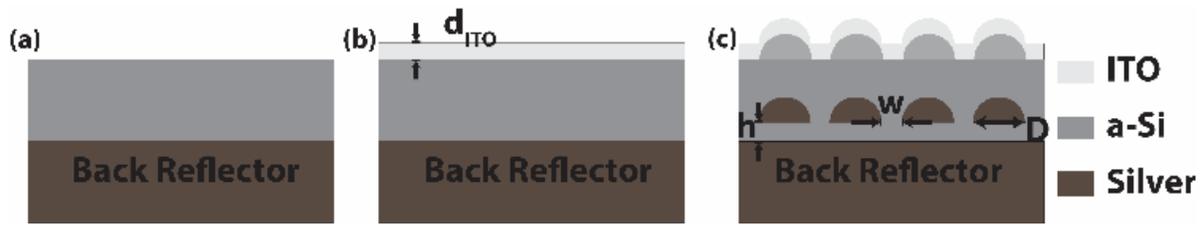

Fig. 1: Cross-sectional view of the device a) bare a-Si with Ag-back reflector, b) ITO/a-Si/Ag-back reflector, c) Hemispherical nano particle for plasmonic behavior (ITO/a-Si (hemispherical Ag nano particle)/Ag-back reflector).

It is well known that beyond 700 nm of solar spectrum, the reflection loss of a a-Si solar cell is very high [4]. Such characteristics is attributed to the low absorption coefficient of a-Si at that said region [4]. Back side nano-structuring is one of the promising technique to trap the region (700 – 800 nm) [6,9,10]. Generally, nano structures scatter light homogeneously when it placed in a homogeneous medium, but the consequence changes when it placed near the interface of two media. Light will then scatter more in higher dielectric medium [8,11]. The scattered light takes an angular spread inside the dielectric medium and increase the path length. Hence, if the nanoparticles are placed in a-Si, the light inside the a-Si will scatter and increase the optical path within it, thus the effective absorption increases. This in-coupling of light in the semiconductor was first observed by Stuart and Hall in 1996 [12]. And in 1998 they have implemented this idea for silicon photodetector [13] and demonstrated 20-fold increase in photo current.

In this work, to maximize the optical absorption at the near infrared region, half spherical nano-structure (HSNS) was used. The diameter (D) and the position of the nano structures form the bottom of the a-Si (h) of the particle was varied from 50–190 and 0–90 nm respectively (shown in : Fig. 1(c)) and further the reflection and absorption of the device was simulated. Such structure can be fabricated by using the techniques described by Nam *et al.*[14]

To compare the performance of different structure of solar cell, the average absorption and reflection of the optical spectrum was determined by using the following relations in the spectral region 350-800 nm,

$$\widetilde{A} = \frac{\int A(\lambda)d\lambda}{\int d\lambda} \qquad (1)$$
$$\widetilde{R} = \frac{\int R(\lambda)d\lambda}{\int d\lambda} \qquad (2)$$

where $\widetilde{A}$ and $\widetilde{R}$ are the average absorption and average reflection. $A(\lambda)$ and $R(\lambda)$ are absolute absorption and absolute reflection at the $\lambda$ respectively.

## 3    Result and Discussion

*3.1  Optimization of ITO thickness*

A bare a-Si solar cell only with the back reflector was simulated to compare the effect of ARC. The reflection and absorption characteristic of a bare a-Si with a back reflector is shown in Fig. 2. It can be observed that the reflection by the a-Si at the low wavelength (<500 nm) region is high which is attributed to the impedance miss-matching. Whereas in higher wavelength (>700 nm) region high reflection fringes is observed which is appearing due to the interference of light from the back reflector and a-Si front surface. This high refection loss will contribute to low absorption in the solar cell and thus low efficiency.

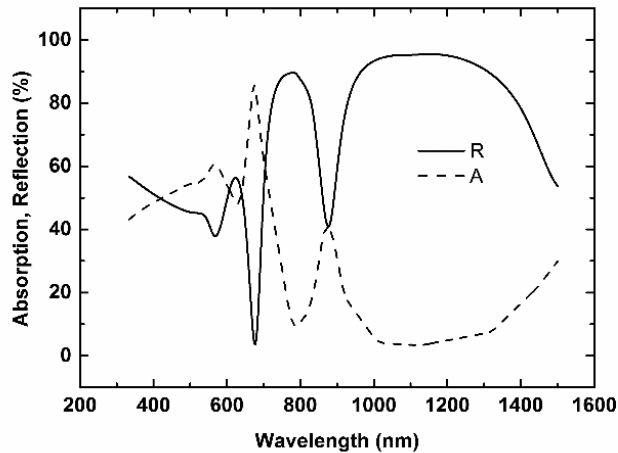

Fig. 2: Reflection and absorption plot of bare-a-Si/Ag-back reflector.

The $\tilde{A}$ and $\tilde{R}$ for the bare silicon was calculated and found to be 47.5% and 52.5% respectively. This observation indicates that the structure has considerable amount of reflection loss which will certainly will degrade the device performance. Therefore, to achieve a better performance, ITO as an ARC layer on the a-Si has to use. The thickness of the ITO plays a crucial role. It works on the principle of quarter wavelength destructive interference. According to the principle, the reflected light and transmitted light interfere destructively and thus it acts like an anti-reflection coating. In order to obtain such condition, the thickness (d) of the anti-reflection coating must satisfy the relation,

$$d = \frac{\lambda}{4n} \qquad (3)$$

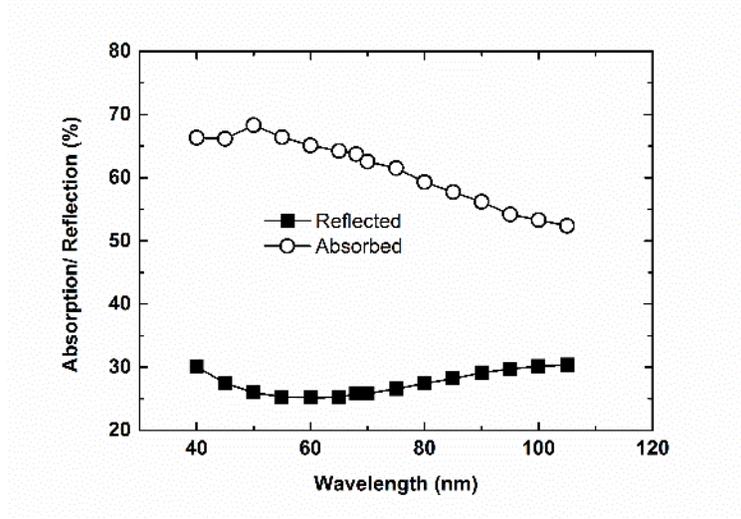

Fig. 3: Optimization of the dITO to obtain maximum average absorption and minimum average reflection. The considered spectrum is 400-800 nm.

Where, d and n (=√(r.i. ambient × r.i. substrate)) is the thickness and refractive index (r.i.) of the ARC layer respectively. Considering the peak emission of solar spectrum at 500 nm ($\lambda$) and n is 1.7 to 2 (for ITO), the thickness should be ~60–65 nm. To optimize more accurate value of '$d_{ITO}$' (thickness of ITO as ARC), the thickness was varried from 40 to 105 nm, and consequently $\tilde{A}$ and $\tilde{R}$ were determined. The plot of optimization is illustrated in the Fig. 3. From the figure, it can be verified that with the thickness of ITO reflection decreases and absorption increases; but, beyond ~65 nm this behavior changes oppositely. The reflection spectrum does not significantly for the thickness 55-70 nm of ITO, therefore for this case a 65 nm of ITO was considered.

The reflection and absorption spectrum of the 65 nm ITO/a-Si is shown in the Fig. 4 (a) and (b) respectively. A comparison with standard 60 nm SiN ARC and bare Si was done. It can be observed that the reflection ($\tilde{R}\sim25\%$) from the structure has reduced effectively and consequently absorption by a-Si ($\tilde{A}\sim66\%$) increases. A slight amount (~9%) of loss is observed which is due to the transmission loss through the ITO. Although use of 60 nm SiN (loss 4%) as an ARC can increase the transmission but ITO is advantageous by its conductive property. Therefore, in this work ITO was used as an alternative ARC material.

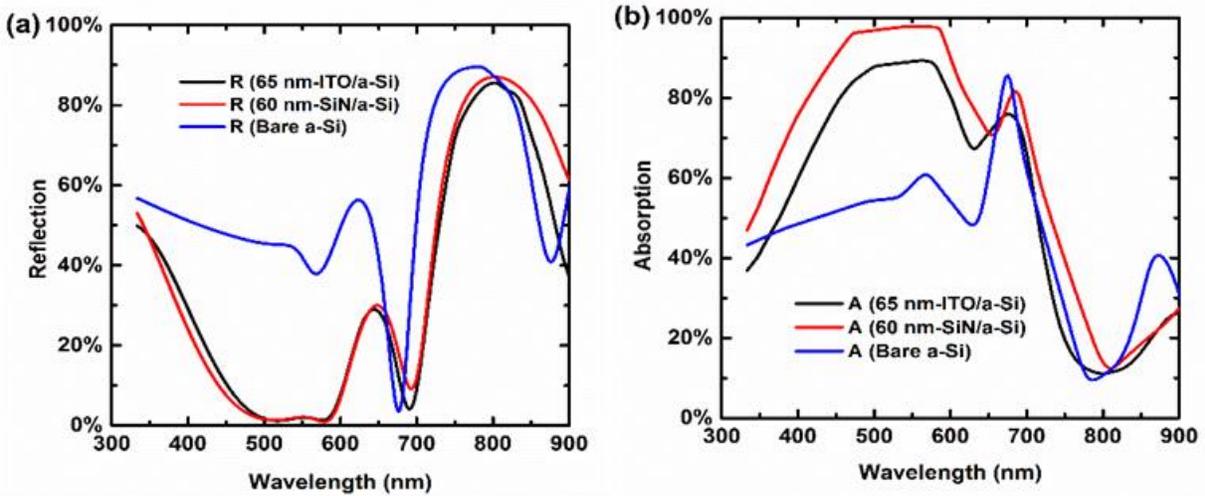

Fig. 4: Comparison of a) reflection and b) the absorption spectrum of the solar cell at different configuration.

*3.2 Nano-structuring and enhancement in performance*

It can be observed form Fig. 4, that beyond 700 nm the reflection from the solar cell increases sharply. This is due to the low absorption by a-Si and followed by the reflection from back metal (since all the structure exhibit similar characteristics). The reflection can be minimized by diverting the light to lateral propagation in the active medium. To achieve that Atwater et. al[8] have used plasmonic nano structure and have demonstrated significant improvement in performance. The nano structures, inserted in semiconductor region scatter the light inside the active medium and increases the path length. Additionally, they generate surface plasmons and stimulate the lateral propagation (or wave guide propagation) of light inside the active medium. Thus, the effective path length increases and consequently the absorption of light enhances. In order to maximize the absorption of light in active medium, the position (h) and the diameter (D) of the HSNS in the active medium were optmized. The cross-sectional schematic of the device is shown in the Fig. 1 (c) (in this case a conformal growth of the thin film was assumed and a slight change in performance has been observed if a combination of conformal and isotropy growth condition will be considered, see supplementary data). During the optimization of D and h, the gap (w) between two nano particle was considered to be 40 nm. The contour mapping of the parameter h and D is shown in the Fig. 5. It can be observed from the Fig. 5(a) that the average reflection ($\tilde{R}$) from the device at the region R1, R2 and R3 is minimum (~6%); and consequently the absorption ($\tilde{A}$) by the a-Si is higher (at R1 and R2), as shown in Fig. 5(b). This attribute to the generation of plasmonic wave at the interface of Si and metal. Therefore, for the simplicity of fabrication process, in this case, region 'R1' was considered as an optimized parameter, and further compared with other devices.

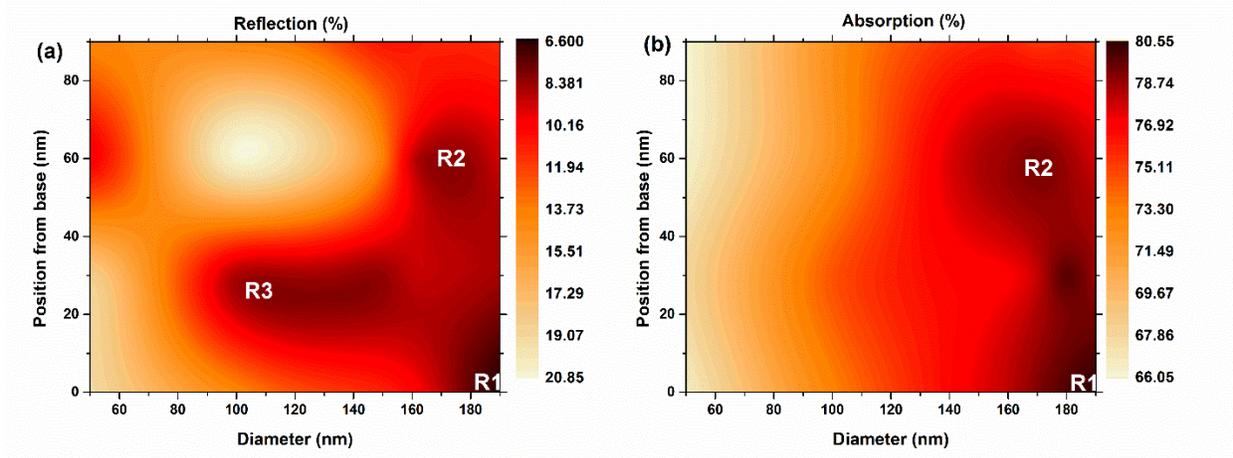

Fig. 5: a) Reflection by the device and b) absorption by a-Si plot to optimize the position and diameter of the nano particle in the solar cell.

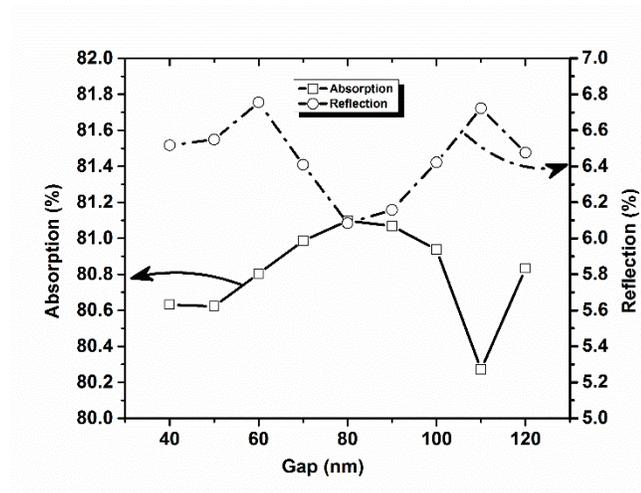

Fig. 6: Optimization of gap (w) between two nano structure, D=190 nm, h=0 nm.

Further, to obtain the maximum absorption by a-Si, the gap (w) between two particles was varied from 40 to 120 nm. The variation of $\tilde{R}$ and $\tilde{A}$ with different 'w' were investigated and shown in the Fig. 6. The figure indicates that the average absorption by a-Si ($\tilde{A}$) is almost constant over the range of w. However, at w=80 nm, the $\tilde{A}$ (~81%) and $\tilde{R}$ (~6%) shows maxima and minima respectively. Therefore, to compare the performance of the nano structure embedded device with the non-structured device, the D and w value is considered to be 190 nm and 80 nm respectively.

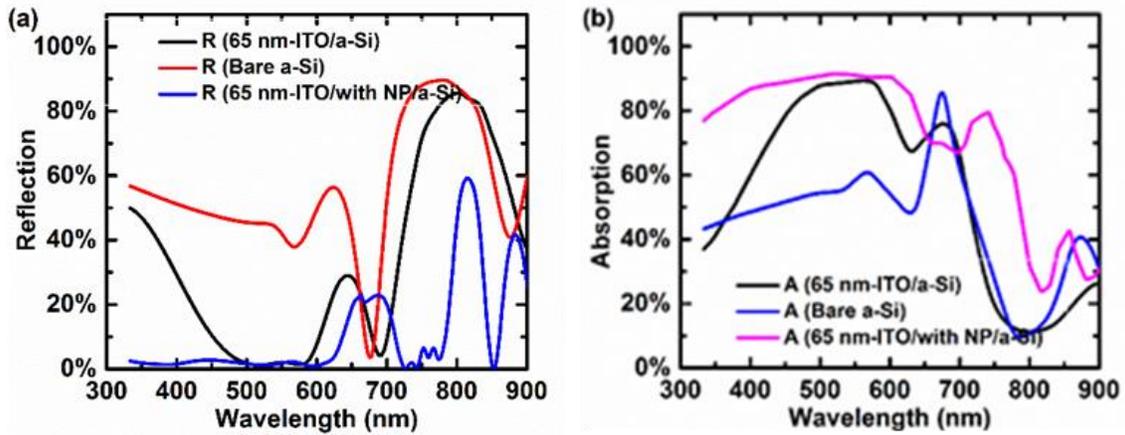

Fig. 7: Comparison of a) reflection spectrum and b) the absorption spectrum of the solar cell at different configuration of solar cell.

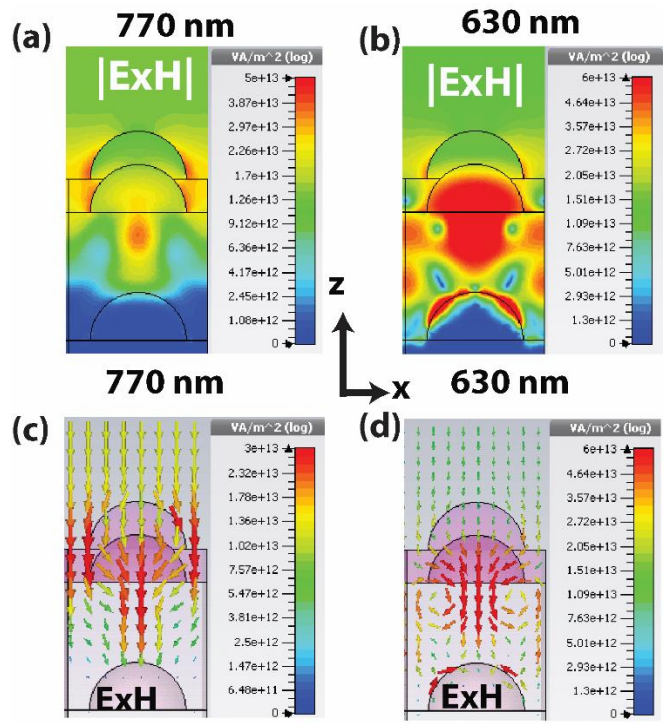

Fig. 8: Simulation results of the power loss in the structure at the (a) wavelength 630 nm and (b) 770 nm. The power flow or Poynting vector direction in the structure at the (c) wavelength 630 nm and (d) 770 nm.

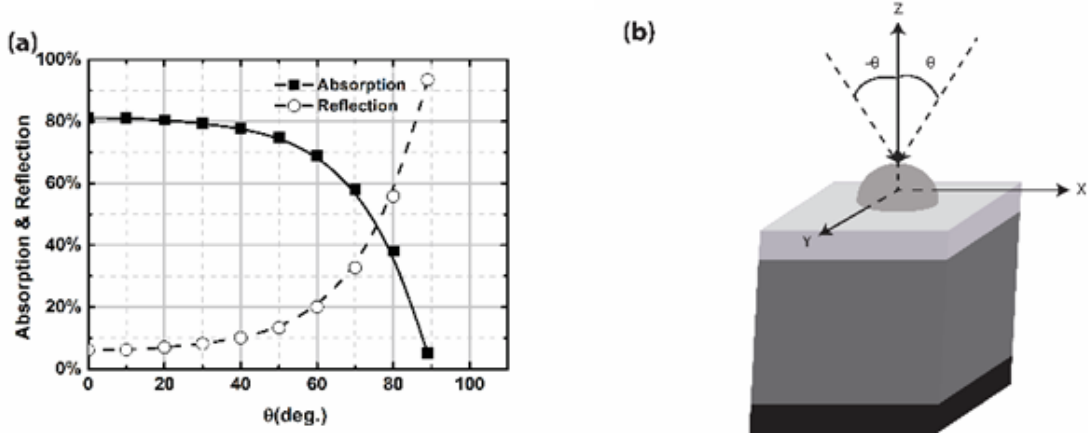

Fig. 9: (a)Absorption and reflection plot in different incident, (b) schematic of 3D diagram of the device with incident light.

The comparison of the reflectance and transmission spectrum of the devices with and without nano particle is shown in Fig. 7. The figure clearly illustrates that the device performance has improved significantly when the nano structured (of size D=190 nm and w=80 nm at h=0) back reflector is used in the device. The enhancement in absorption can be explained by generation of surface plasmon by the HSNS. It is evident from the Fig. 8 (c)&(d), the power flow direction (Poynting vector) of the propagating wave (shown in the Fig. 8) is tangential inside the a-Si. The incident wave propagation vector of the light is considered along the z-axis, which can be observe at outside of the structure (in Fig. 8 (c) & (d)). But as the light enters in the structure it starts lateral propagation, this results to increases in optical path in a-Si. And thus the optical absorption by a-Si increases effectively. Although, a small degradation in absorption in the visible region (657-694 nm) is observed, but spectral absorption enhances significantly (~80%). The optical power loss in the structure is shown in the Fig. 8 (a) and (b), which shows that most of the power is absorbed in a-Si region. This characteristic is required to enhance the photo carrier generation and the quantum efficiency of the device.

Finally, the performance the solar cell was investigated under oblique incidence of light to study the angular coverage of solar cell. The angle was varied with respect to normal of the solar plane (with respected to the z-axis as shown in the figure 9 (d)). The performance of the cell was calculated by irradiating a fixed amount of light to the cell. It can be observed from Fig. 9 (a) that the absorption of the solar cell is significant (varies 81% to 60%) for the incident angle of 0 to 70 deg and decrease sharply after that. This characteristic indicates that the proposed structure has large angular coverage. So, it has been verified that the ITO ARC along with nano back structure can enhance the optical absorption in a-Si thin film solar cell in the visible and near infrared region with large angular coverage. Therefore, proposed structure will improve the quantum efficiency of the solar cell.

## 4. Conclusion

In this paper, ITO was used as an alternative ARC for a-Si thin film solar cell in order to obtain a broad band anti-reflection property in the visible and near infrared region (350 – 800 nm) of spectrum. The thickness of the ITO was optimized and it was found to be 65 nm. By using ITO, the reflection from the structure has been reduced significantly to 22% from 52%. And consequently, the absorption of a-Si has been increased to 68% from 47%. Furthermore, in order to enhance the absorption at the band edge of a-Si the nano structured back reflector was used. And it was found that the spectral absorption (in 350–800 nm) has improved to ~80% when and the reflection loss minimized to 6%. The angular span of the device is found to be promising. Therefore, by using ITO and nano structured back reflector the device performance can increase effectively.

**Acknowledgement**

The authors would like to thank Asian Office of Aerospace Research and Development (US Army) for the funding of the project.

**Funding Information:** Asian Office of Aerospace Research and Development (US Army), Project: 14AOARD001.